\documentclass[journal]{IEEEtran}
\usepackage{amsmath}
\usepackage{graphicx}

\usepackage[dvips]{color}
\usepackage{color}
\usepackage{cite}

\newcommand{\minorchange}[1]{#1}

\title{Comparing Maintenance Strategies for Overlays
\thanks{This work is funded by the 6th FP EVERGROW project.}
}
\author{
          Supriya Krishnamurthy$^{1,3}$, Sameh El-Ansary$^{1}$, Erik Aurell$^{1,2}$ and Seif Haridi$^{1,3}$\\
          $^1$ Swedish Institute of Computer Science (SICS), Sweden\\
          $^2$ Department of Computational Biology, KTH-Royal Institute of Technology, Sweden\\
					$^3$ IMIT, KTH-Royal Institute of Technology, Sweden\\
        	\{supriya,sameh,eaurell,seif\}@sics.se \\     	
}

\begin{document}
\maketitle

\begin{abstract}
In this paper, we present an analytical tool for understanding 
the performance of structured overlay networks under churn based on 
the master-equation approach of physics. We motivate and derive
an equation for the average number of hops taken by lookups
during churn, for the Chord network.
We analyse this equation in detail to understand 
the behaviour with and without churn. We then use this
understanding to predict how lookups will scale for
varying peer population as well as varying the sizes of the 
routing tables. We then consider a change in the maintenance
algorithm of the overlay, from periodic stabilisation to
a reactive one which corrects fingers only  when a 
change is detected. We generalise our
earlier analysis to understand how the reactive strategy compares with
the periodic one.

\end{abstract}

\section{Introduction}
\label{intro}
A crucial part of assessing the performance of a structured P2P system (aka DHT) is
evaluating how it copes with churn. Extensive simulation is currently the prevalent tool for 
gaining such knowledge. Examples include the work of Li \textit{et al.} \cite{dhtcomparison:infocom05}, 
Rhea \textit{et al.} \cite{rhea04handling}, and Rowstron \textit{et al.} \cite{rowstron04depend}. There has also been some theoretical analyses done,  albeit less frequently. For instance, Liben-Nowell \textit{et al.} \cite{nowell02analysis} prove a lower bound on the 
maintenance rate required for a network to remain connected 
in the face of a given churn rate. Aspnes  \textit{et al.} \cite{aspnes02FaultTolerant} give upper and lower 
bounds on the number of messages needed to locate  a node/data item 
in a DHT in the presence of node or link failures. 
The value of theoretical studies of this nature is that they provide 
insights neutral to the details of any particular DHT.  

We have chosen to adopt a slightly different approach to theoretical 
work on DHTs. We concentrate not on establishing bounds, but rather on
a more precise prediction of the relevant quantities in such 
dynamically evolving systems. Our approach is based mainly on the Master-Equation approach used in the
analysis of physical systems. We have previously introduced our approach in 
in \cite{KEAH1,KEAH2} where we presented 
a detailed analysis of the Chord system \cite{chord:ton}. 
In this paper, we show that the approach is applicable to other systems as well.
We do this by comparing the periodic stabilization maintenance technique 
of Chord with the
correction-on-change maintenance technique of DKS \cite{onana03dks}.  

Due to space limitations, we assume reader familiarity 
with Chord and DKS, including such terminology as successors, finger starts 
and  finger nodes {\it etc}.

The rest of the paper is organised as follows. In Section \ref{mastereq}, we  introduce the Master-Equation 
approach. In Section \ref{related}, we mention some related work. In section \ref{lookup} we begin by briefly 
reviewing some of
our previously published results on predicting the performance of the Chord network
as a function of the failed pointers in the system in the case that the nodes
use a periodic maintenance scheme. We then show some new results on how this complicated
equation can be simplified to get quick predictions for varying number
of peers and varying number of links per node. We relegate some of the details of 
this analysis to Appendix \ref{A1}. In section \ref{correction-on-change}, 
we explain how to use the Master-Equation approach to analyse the 
reactive maintenance strategy of interest and present 
our results on how this strategy
compares with the periodic case analysed earlier. We summarise our results 
in Section  \ref{summary}.

%Also something about the rates (failure, join and stabilisation, $alpha$, $r$) for the
%periodic case. The rates used in the reactive case are explained in the 
%appropriate section.

\section{The Master-Equation Approach for Structured Overlays}
\label{mastereq}
In a complicated system like a P2P network, in which 
there are many participants, 
and in which there are many inter-leaved processes happening in time, 
predicting the state of the network (or of any quantity of interest) can 
at best be done by specifying the 
probability distribution function (PDF) of the 
quantity in the steady state  (when the system, though changing 
continually in time, is stationary on average).
For example, one quantity of interest for us when analysing such a network, is the fraction of failed links between nodes, in the steady state. 
This quantity does not take 
some deterministic value in the steady state. Instead it is specified by a PDF,
which can then be used to determine the average value.
The problem is thus to calculate the PDF (and then to understand how it
affects the performance of the network, as explained below).

In general this is not
an easy task, since the probability is affected by a number of inter-leaved 
processes in any time-varying system.  In \cite{KEAH1,KEAH2}, we demonstrated 
how we could analyse a P2P network like Chord
\cite{chord:ton},  using a Master-Equation based approach.  This approach
is generally used in physics to understand a system evolving in time,
by means of equations specifying the time-evolution of the probabilities of finding the 
system in a specific state. These equations require as an input, 
the rates of various processes affecting the state of the system. For example, 
in a peer-to-peer network, these processes could be the join and failure rates 
of the member nodes, the rate at which each node performs maintenance as well as 
the rate at which lookups are done in the network 
(the latter rate is relevant only if the lookups affect the state of the 
network in some way). Given these rates, the equation for the time-evolution 
of the probability of the quantity of interest can be written by
keeping track of how these rates affect this quantity (such as the number 
of failed pointers in the system) in an infinitesimal interval of time, 
when only a limited number of processes (typically one) can be expected to 
occur simultaneously.

 With this approach, we
were able to quantify very accurately the probabilities of any connection in the network 
(either fingers or successors) having failed. We then 
demonstrated how we could use this information 
to predict the performance of the network---the number of hops {\it including} time outs
which a lookup takes on average --- as a function  of the rates 
(of join, failure and stabilization) of all the processes happening in the network, as
well as of all the parameters specifying the network (such as
how many pointers a node has on average). The analysis was done for a specific 
maintenance strategy, called periodic maintenance (or eager maintenance)

In this paper, we generalise our approach so as to be able to compare networks
using different maintenance strategies. In particular, we compare our earlier results 
for periodic maintenance with a reactive maintenance strategy proposed in
\cite{ghodsi05hicss}. Combining this with some of our previous results, we are
also, as a by product, able to compare the performance of networks specified by 
different  numbers of peers, different number of pointers per node and/or 
different maintenance strategies. As we show below, which system is better
depends both on the value of the parameters as well as the level of churn. 
The approach we propose is thus a useful tool for the quantitative and fair 
comparison of networks specified by different parameters and using different algorithms.

\section{Related Work}
\label{related}
In \cite{aberer04tok}, an analysis, very similar in spirit to the one done in this paper, is carried out in
the context of P-Grid \cite{pgrid}. An equation is written for system performance in the 
state of dynamic equilibrium for various maintenance strategies. However for each maintenance strategy, 
the analysis has to be entirely redone. In contrast, a master equation description 
provides a foundation for the theoretical analysis of overlays, 
which does not have to be entirely rebuilt each time any given algorithm is changed. 
As we show in this paper, we can carry over a lot of our earlier analysis, when the maintenance scheme
is changed from a periodic to a reactive one. In addition, the master equation description
can be made arbitrarily precise to include non-linear effects as well. And as we show, non linear effects 
are important when churn is high.

\section{The Lookup Equation for Chord}
\label{lookup}

We quantify the performance of the network, by the number of hops required 
on average from the originator of the query to the node with the answer. This 
is just the total number of nodes contacted per query 
(or equivalently, the total
number of pointers used per query) {\it including }
the total number of failed pointers used
en route. This latter quantity 
(which arises because of the churn in the network)
is the reason that the hop count per query increases with high dynamism and is
hence an important quantity to understand. In the case
of the periodic maintenance scheme, this quantity is a function of
$(1- \beta) r$ where $r$ is the ratio of the stabilisation rate to the 
join (or failure) rate and $1-\beta$
is the fraction of times a node stabilises its finger,  when 
performing maintenance, as mentioned in Section \ref{intro}.
We demonstrate how this quantity can be calculated in Section 
\ref{correction-on-change}, in the context of the reactive 
maintenance policy,
which is a simple generalisation of how it is calculated 
earlier in\cite {KEAH1,KEAH2      }, 
for the periodic maintenance scheme.
In this section, we briefly review our earlier results on  how the 
performance of the network (as exemplified by the average hopcount per query), 
can be determined once the fraction of failed pointers is known.

The key to predicting the performance of the network is to 
write a recursive equation
for the expected cost  $C_{t}(r, \beta)$ 
(also denoted $C_{t}$) for a given node 
to reach some target, $t$ keys away from it.
(For example,  $C_1$ is the cost of looking up the adjacent key which is $1$
key away).

The Lookup Equation for the expected cost of reaching a general distance $t$
is then derived by following closely the  Chord 
protocol which is a greedy strategy
designed to reduce the distance to the query at every step 
without overshooting the target .  A lookup for $t$ 
thus proceeds by first finding 
the closest preceding finger. The node that this finger points 
to is then asked to continue the query, if it is alive. 
If this node is dead, the originator of the query
uses the next closest preceding finger and the query proceeds in this manner.

For the purposes of the analysis,
it is easier to think in terms of the closest preceding {\it start}.
Let us hence define $\xi$ to be the {\emph start} of the 
finger (say the $k^{th}$) that most closely precedes $t$. 
Hence $\xi = 2^{k-1} + n$ and
$t = \xi+m$, i.e. there are $m$ keys between the sought target $t$ 
and the start of the most closely preceding
finger.  With that, we can write a recursion relation 
for $C_{\xi+m}$ as follows:

\begin{equation}
%\vspace*{-0.5cm}
\label{eq:cost}
\begin{split}
&C_{\xi+m} =  C_{\xi} \left[1-a(m)\right]  						\\
				         &+ (1-f_k) a(m)\left[1 + \sum_{i=0}^{m-1} bc(i,m)C_{m-i}\right]					
          \\					 
					 &+ f_k  a(m) \biggl[ 1 + \sum_{i=1}^{k-1} h_k(i) \\
					 &\sum_{l=0}^{\xi/2^i-1}bc(l,\xi/2^i)(1+(i-1) +C_{\xi_i-l+m}) + O(h_k(k)) \biggr]
%					 &+ \biggl[ f_k  a(m) \\
%					 &+f_k  a(m)\sum_{i=1}^{k-1} h_k(i) \sum_{l=1}^{\xi/2^i}bc(l,m)(1+C_{\xi_i(k)+1-l+m}) + 2h_k(k)\biggr] 
\end{split}			
%\vspace*{-0.5cm}		
\end{equation}

where $\xi_i \equiv \sum_{m=1,i} \xi/2^{m}$ and $h_k(i)$ is the 
probability that a node is forced to use its $k-i^{th}$ finger owing to the 
death of its $k^{th}$ finger.

The probabilities $a,bc$  can be derived from the internode 
interval distribution
\cite{KEAH1,KEAH2} which is just the distribution of distances between adjacent nodes. Given a ring of $\cal K$ keys and $N$ nodes (on average), where nodes can join and leave independently, the probability that two adjacent nodes are a distance $x$ apart on the ring is simply
$P(x) = \rho^{x-1}(1-\rho)$ where $\rho = \frac{{\cal K}-N}{\cal K}$.
Using this distribution, its easy to estimate the probability that 
there is definitely atleast one node in an interval of length $x$.
This is: $a(x)\equiv {1-\rho^x}$. The probability 
that the {\it first} node encountered from any key is at a distance $i$ from that key is  then $b_i \equiv {\rho^{i}(1-\rho)}$. 
Hence the conditional probability that the first node from a given key is at a distance $i$ {\it given} that
there is atleast one node in the interval is $ bc(i,x)\equiv b(i)/a(x)$.

The probability $h_k(i)$ is easy to compute given the probability $a$ 
as well as the probabilities  $f_k$'s of the $k^{th}$ finger being dead.

\begin{equation}
\label{eq:hki}
\begin{split}
h_k(i) = & a(\xi/2^{i}) (1-f_{k-i})  \\
       \times &\Pi_{s=1,i-1} (1-a(\xi/2^{s}) + a(\xi/2^s)f_{k-s}), i<k \\
h_k(k) = & \Pi_{s=1,k-1} (1-a(\xi/2^{s}) + a(\xi/2^s)f_{k-s})  
\end{split}
\end{equation}

Eqn.\ref{eq:hki} accounts for all the
reasons that a  node may have to use its $k-i^{th}$ finger 
instead of its $k^{th}$ finger. This could happen because the 
intervening fingers were either dead or not distinct (fingers $k$ and $k-1$ are not distinct if 
they have the same entry in the finger table. Though the {\it starts} of the two fingers are
different, if there is no node in the interval between the {\it starts}, the entry in the finger table will be the same).
The probabilities $h_k(i)$ satisfy the constraint $\sum_{i=1}^{k} h_k(i)=1$.
$h_k(k)$, is the probability that a node
cannot use any earlier entry in its finger table,in which case it has to fall back on its successor list instead.
We indicate this case by the last term in Eq. \ref{eq:cost} which is 
$O(h_k(k))$. In practise, the probability for this
is extremely small except for targets very close to $n$. 
Hence this does not
significantly affect the value of general lookups and we ignore it
for the moment.

The cost for general lookups is 
$$
L(r,\beta) = \frac{\Sigma_{i=1}^{{\cal K} -1} C_i(r,\beta)}{\cal K} 
$$

\begin{figure}[t]
	\centering
		\includegraphics[height=8cm, angle=270]{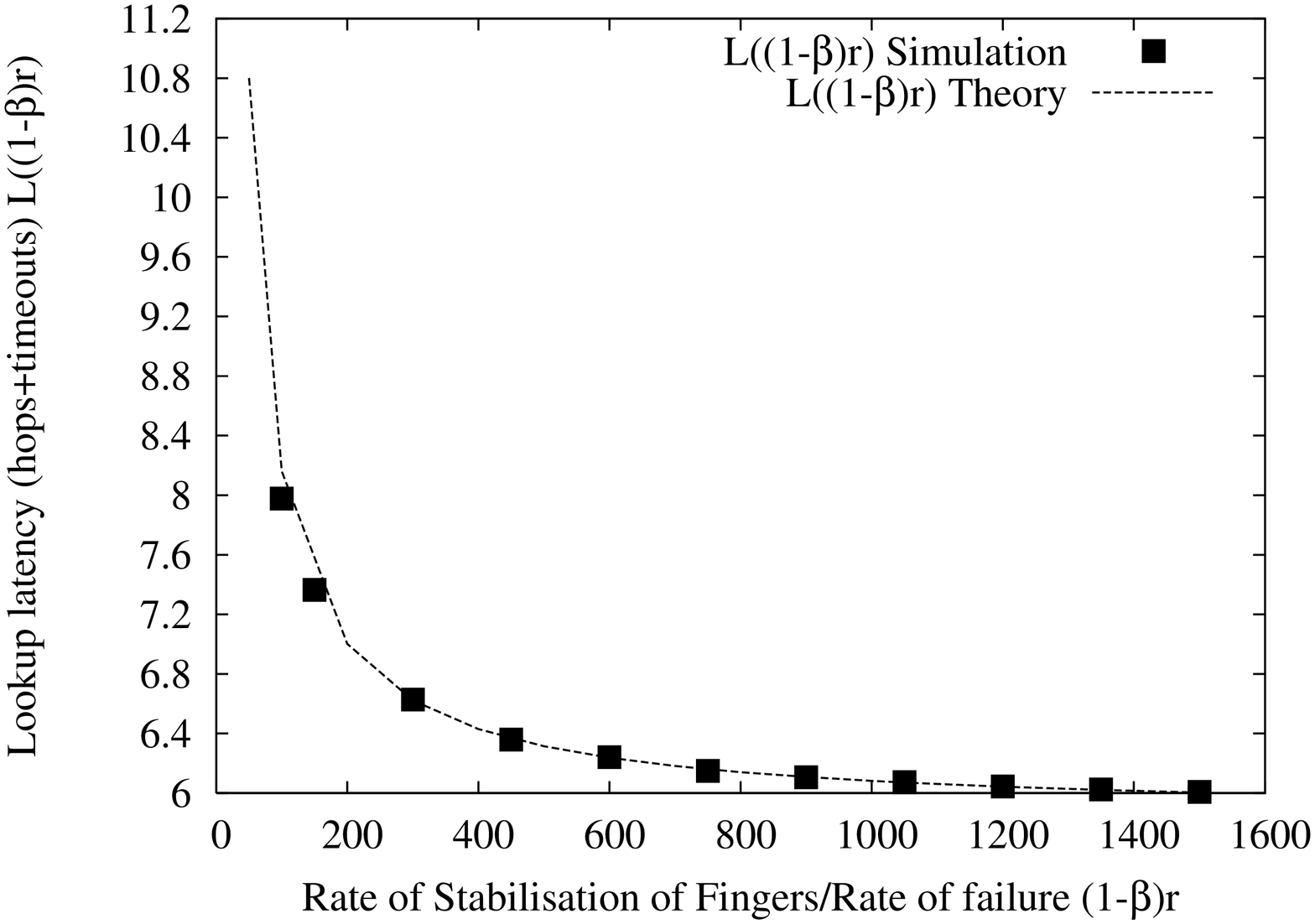}

		%\begin{table}[t]
	   %\centering
%	\vspace*{-0.3cm}
	\caption{Theory and Simulation  for $L(r,\beta)$}
	\label{fig:w}
\end{figure}

The lookup equation is solved recursively numerically, using the expressions for 
$a$, $bc$, $h_k (i)$ and $C_1$. In Fig. \ref{fig:w}, 
we have plotted the theoretical prediction of Equation \ref{eq:cost}
versus what we get from simulating Chord. Here we have used 
$N \sim 1000$ and $K= 2^{20}$. As can be seen the 
the theoretical results match the simulation results very well.

\begin{figure}[t]
	\centering
	\includegraphics[height=8cm, angle=270]{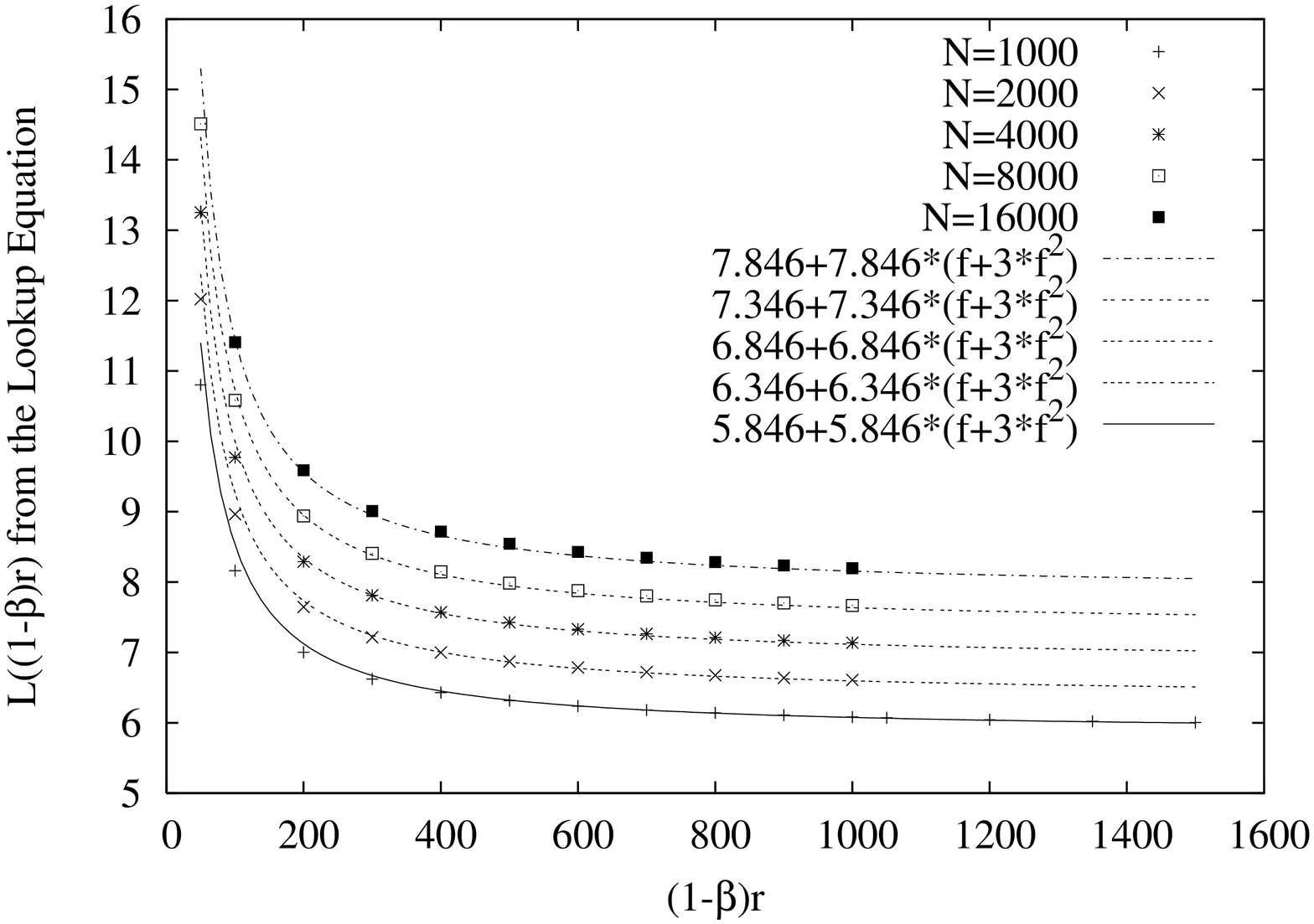}
	\caption{Lookup cost, theoretical curve, for $N=1000, 2000,4000,8000, 160000$ peers. The rationale for the fits is explained later in the text.}
	\label{fig:lookup_theory}
\end{figure}

In Fig.~\ref{fig:lookup_theory} we also show the theoretical predictions
for some larger values of $N$.

On general grounds, it is easy to argue from the structure of 
Equation \ref{eq:cost}, that the dependence of the average lookup 
on churn comes entirely 
from the presence of the terms $f_k$. 
Since $f_k \sim f$ is independent of $k$ for large fingers, we can
approximate the average lookup length by the  
functional form $L (r, \beta) = A + {B}f + C f^{2} + \cdots $.
The coefficients $A, B, C$ {\it etc} can be recursively computed 
by solving the lookup equation
to the required order in $f$. They depend only on $N$ the number of nodes, 
$1- \rho $ the  
density of peers and $b$ the base or equivalently the 
size of the finger table of each node. 
The advantage of writing the lookup length this 
way is that churn-specific details such as how new 
joinees construct a finger table 
or how exactly
stabilizations are done in the system, can be isolated 
in the expression for $f$.
If we were to change our stabilization strategy, as we will
demonstrate below, we could immediately
estimate the lookup length by plugging in the new expression 
for $f$ in the above relation.

Another advantage of having a simple expression such as the above, 
is that if we can estimate $A,B, C \cdots$ accurately, we can make use
of the expression for $L$ to estimate the churn 
(or the value of $r$) in the system, 
hence using a local measure to estimate a global quantity.
The logic in doing so is the inverse of the reasoning we have used  so far.
So far, we have used the churn 
 as the input for finding $f_k$ and hence $L$. But we can also reverse 
the logic and try and estimate churn, if we know the value of the 
average lookup length $L$. If $L$ has the above simple expression, 
then given $A$ and $B$ to $O(f)$, we have
$f= \frac{L-A}{B}$. From  the expression for $f$ 
(see section \ref{correction-on-change} for how to evaluate $f$), 
we can now get the value of $r$. Hence any 
peer can make an estimate of the churn that the system 
is facing if it knows how long its lookups are taking on average, and 
if it has an estimate of $N$.

To get $A$, we need to consider Eqn \ref{eq:cost} with no churn 
(all $f_k$'s set to zero). 
%To get $B$, we  need to analyze the lookup equation to $O(f)$ and so on.
In Appendix \ref{A1}, we study 
the lookup equation ~\ref{eq:cost} in some detail to understand
the behaviour without churn and obtain the value of $A$ for any base $b$.
This is useful on several counts. First,
the value of $A$ is needed to predict the lookup costs as explained above.
Secondly, if $b$ changes ( a system of base $b$ has a finger table of size
${\cal M}=(b-1)log_b ({\cal K})$), all else remaining the same, the only 
major change in the lookup cost is due to the change in $A$. So estimating
$A$ precisely has the benefit that we 
can predict the lookup cost for {\it any} 
base $b$. Thirdly, the analysis confirms that Equation \ref{eq:cost}
does indeed reproduce well known results for 
the lookup hop count in Chord, such
as for example, that the average lookup cost is $0.5*\log(N)$
without churn \cite{chord:ton}. Infact as demonstrated in Appendix \ref{A1},
for any $N$, the average lookup cost as predicted by Eq. \ref{eq:cost} 
is indeed $0.5*\log(N)$ plus some $\rho$-dependent corrections which 
though small are accurately predicted. 

%An added benefit of the analysis is
%that we can also predict what the average lookup without churn will be for
%any base (Chord has base $2$ and accordingly has a finger table 
%size of $log_2({\cal K})$. 

A simple estimate for $B$ and $C$ can be made in the following manner. 
Let every finger be dead with some finite probability $f$. 
Each lookup encounters on average $A$ fingers,
where $A$ is the average lookup length {\it without} churn. 
Each of these fingers could be alive 
(in which case it contributes a cost of $1$), 
dead with a probability $f$ in which case it contributes a cost of $2$ if the
next finger chosen is alive (with probability $1-f$) and so on. Its trivial to 
verify that this estimates the look-up cost to be $A(1+f + f^2 + \cdots)$.
Comparing with our expression for $L$, this gives an estimate of 
$B= A, C=A, \cdots$. 

\begin{figure}[t]
	\centering
	\includegraphics[height=8cm, angle=270]{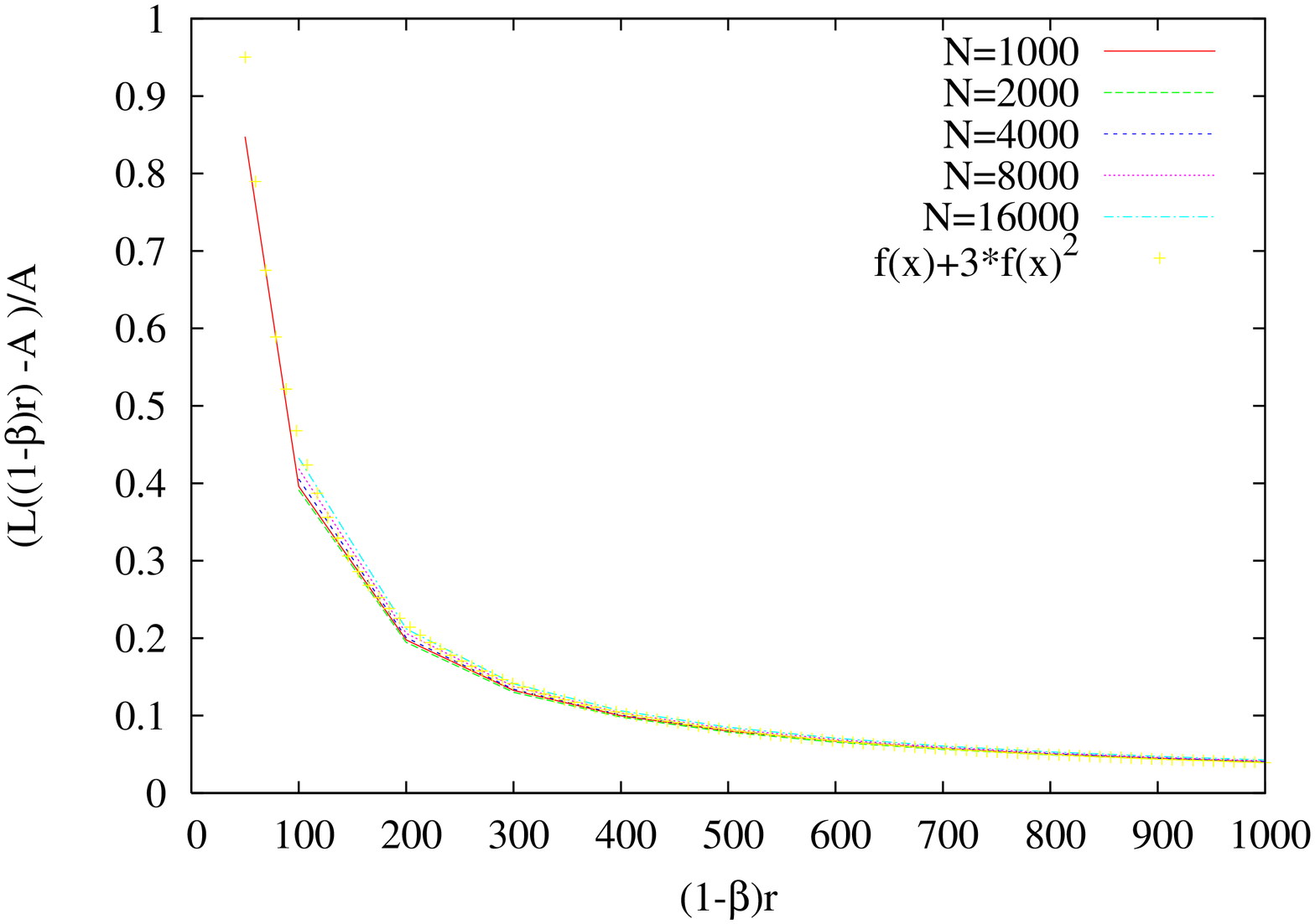}
	\caption{Scaled Lookup cost, for $N=1000, 2000,4000,8000, 160000$ peers.}
	\label{fig:lookup_theory_scaled}
\end{figure}

In general if $L=A+B*g(f)$,
then if we scale $L$ by plotting $(L-A)/B$ for varying $N$, 
we should get an estimate of $g(f)$. Note that $f$ depends on $\rho$ and 
${\cal M}$ the number of fingers. In addition if $g(f)= a_1f+a_2f^2 + \cdots$, the coefficients $a_1$,$a_2$, {\it etc} 
can also depend on $\rho$.
However for $1-\rho <<1$, these dependences on $\rho$ are small and
the curves for different $N$ collapse onto the same curve on scaling.
In Fig. ~\ref{fig:lookup_theory_scaled} we have scaled the curves ploted 
in Fig. ~\ref{fig:lookup_theory} in the above manner, using $B=A$. 
The values of $A$ used are derived from the analysis of the previous section. 
As can be seen the curves collapse onto one curve which is well 
approximated by the function $g(f)=f+3*f^2$, giving $a_1=1$ and $a_2=3$.
The fits in Fig ~\ref{fig:lookup_theory} are also according to this functional form. 
It should be emphasized however that this approximation for $g(f)$ is 
good only for $1-\rho <<1$. For higher values of peer density, the curves for
different $N$ will not collapse onto one curve and 
any $\rho$-dependence of the coefficients $a_i$'s will show up as well.

\begin{figure}[t]
	\centering
	\includegraphics[height=8cm, angle=270]{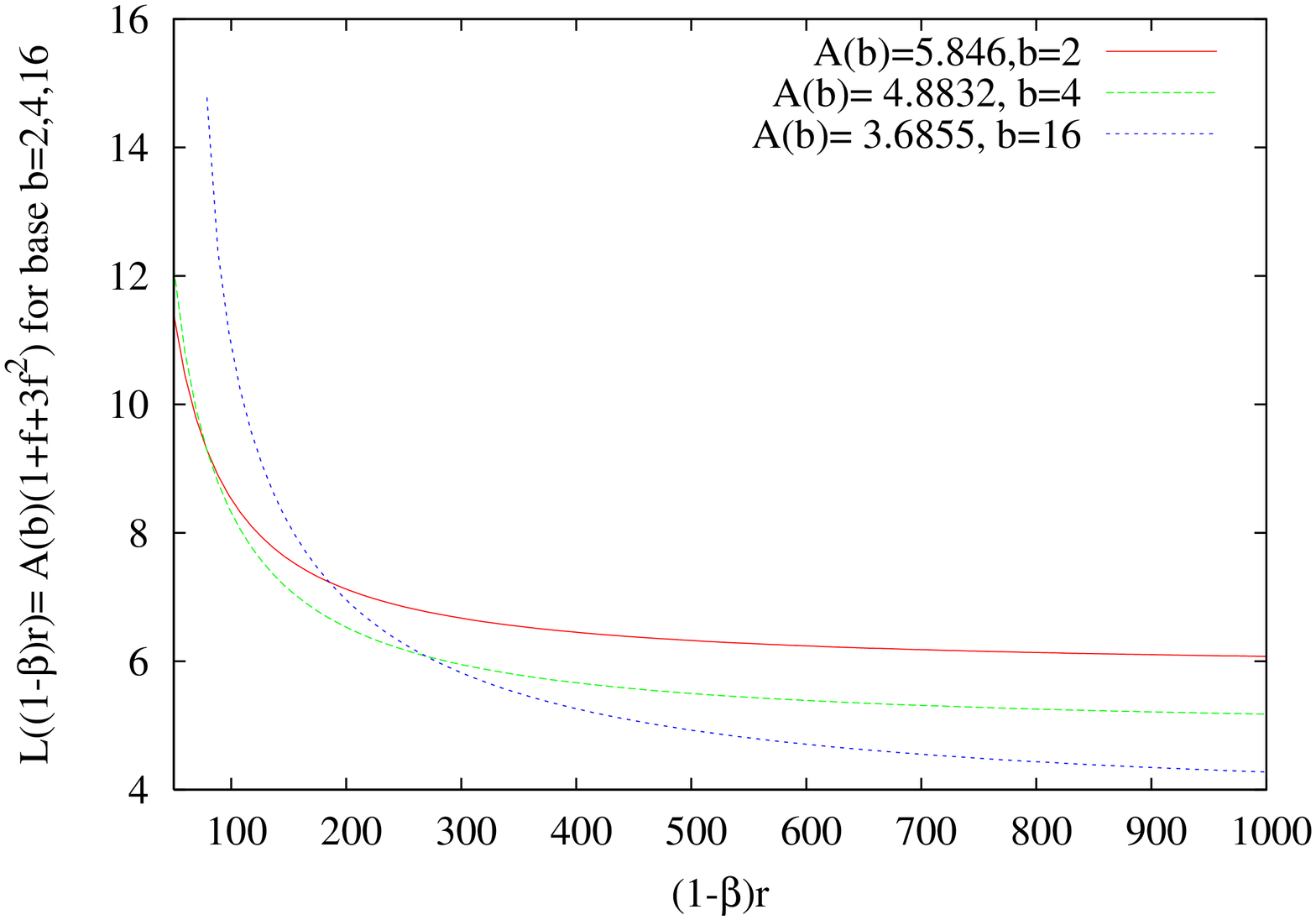}
	\caption{Lookup cost, for $N=1000$ peers for base $b=2,4,16$.}
	\label{fig:lookup_theory_base}
\end{figure}

We can use the above functional form to predict how lookups would behave if we change the base 
$b$ (the size of the routing table) of the system. In Fig 
~\ref{fig:lookup_theory_base} we plot the functional form
$A(b)(1+f(b)+3f(b)^2)$ for $b=2,4,16$. The coefficient $A(b)$ is
accurately predicted by Eq. \ref{eq:constnochurn1}(in Appendix \ref{A1}), 
with the definition of $\xi(i+1)$ taken appropriately. $f(b)$ is affected by the base $b$ 
because the number of fingers increases with $b$.

As can be seen, when churn is low, a large $b$ is an advantage and 
significantly improves the lookup length. However when churn is high, 
the flip side of having a larger routing table is that it needs 
more maintenance. Hence beyond some value of churn, the
larger the value of $b$, the larger the lookup latency. 

This is similar to the spirit of the numerical investigations done in ~\cite{li03comparing}.
However when comparing different bases for Chord, Li {\it et al} ~\cite{li03comparing} find
that while base $2$ is the best for high churn (as we find here), base $8$ is the best for low churn. 
Increasing the base beyond this does not seem to improve the cost. The discrepancy between this finding 
and ours is due to the details of the periodic maintenance scheme which we use. 
In our case, we have taken the simplest scenario in which each node needs to stabilise ${\cal M}$ fingers
and the order in which this is done is random. In practice 
only $\sim \log N$ of the ${\cal M}$ fingers are distinct, so only $\sim \log N$ stabilisations need be done 
by each node. In addition, in  ~\cite{li03comparing}, finger stabilisations are done {\it only} if the
finger is pinged and found to be dead. 

\section{'Correction-on-Change' Maintenance Strategy}
\label{correction-on-change}

In this section, we analyse a different maintenance strategy 
using the master-equation formalism. The strategy we have analysed 
so far is periodic stabilisation of successors as well as fingers.
We now consider a strategy where a node periodically stabilises its 
successors but does not do so for its fingers. Instead, for maintaining 
its fingers, it relies on other nodes for updates \cite{ghodsi05hicss}.
Whenever a node $n$ detects that its first successor $n.s_1$ is wrong 
(failed or  incorrect), it sends out messages to all the nodes 
that are pointing 
to its  wrong first successor, so that they can update 
their affected finger. The node sending messages can either do so
by broadcasting these messages to all affected nodes simultaneously, or by
scheduling messages periodically at some rate. We analyse the latter option
in this paper, since it provides a more intuitive and broader framework for
the comparison of the two schemes

For a system with {\it id}-size ${\cal K}$,
there are of the order of ${\cal M}=\log_{2}{\cal K}$ fingers 
pointing to any node (there can be more than this if node spacings are
smaller than average. However, as we argue below, for our purpose this is not important).
Of course, not all $\cal M$ of these fingers are distinct. Several of these 
fingers belong to node $n$ itself.
However to keep the analysis simple (and in keeping with the 
spirit of our analysis of the periodic stabilisation scheme), we assume that
every node that detects a wrong successor needs to send out exactly 
$\cal M$ messages (even if some of these 'messages' are sent to itself).

To find out where the nodes that point to $n.s_1$ are located, $n$ 
needs to do a lookup. For example, to find the node 
with the $k^{th}$ finger pointing to $n.s_1$, $n$ can do a lookup for the id $n-2^{k-1}$. 
On obtaining the first successor (lets call it node $p$)
of this id, it would immediately know
if the $k^{th}$ finger of $p$ indeed needs to be updated. We think
of each lookup as a 'correction message'.
If there is  more than one node that needs its $k^{th}$ successor updated
(because for example, the successors of $p$ also happen to point to $n.s_1$), 
$n$ could leave the responsibility of informing these other nodes to 
$p$. We could take into account the probability that a correction action
leads to more than $\cal M$ messages. But for the moment we ignore 
this point (We could argue that once it is $p$'s responsibilities to check
that its successors know about $n.s_1$, it could piggy-back this information 
when it does a successor stabilisation, which does not affect the number
of messages sent).

Whenever a node receives a message updating its 
information about a finger, it immediately corrects the appropriate entry 
in its routing table.

In the following, we demonstrate how we can analyse such a strategy. 
We would like to ultimately compare its performance to
periodic stabilisation in the face of churn. To make such a comparisn 
meaningful, we need to quantify the concept of 'maintenance-effort' 
per node, and compare the two schemes at a given level of churn and at the same
value of the maintenance effort per node.We elaborate on this a little later
in Section \ref{compare}.

%A simple way is to look at it as just the total number of 
%messages sent per node. Here we
%are simply adding up all the successor-stabilisation actions performed 
%by the node as well as all the correction messages sent, 
%in order to get the total number of messages sent per node.
%This is not strictly correct since a correction-message takes of 
%the order of $\log_2(N)$ hops while a successor-stabilisation takes 
%only one. But we keep it this way for the sake of simplicity
%(and also because we have analysed periodic stabilistion this way).
%To be more accurate, we may have to take into account the 
%different natures of successor stabilisation and correction messages.
%We may also need to take into account the amount of {\it information} 
%sent per message instead of just the total number of messages.
%However in the following, we take the simplest route.

%Another complication here is that in the case of periodic stabilisation,
%successor and finger stabilisations are independent and can happen
%at independent rates. However for the 'correction-on-change' strategy
%described above, the rate at which correction-messages are sent, 
%is dependent on the rate at which successor stabilisation is done. 
%We will quantify this below.

Another point to note is how to quantify system performance. We have 
previously done it in terms of lookup hops. But a more
correct way might be to ask for the latency for {\it consistent} lookups
(since some of the lookups could be inconsistent). However we have checked that
, within our analytical framework, this does 
not change the results qualiltatively.

\subsection{Analysis of the Correction-on-change strategy }
\label{analysis}

To generalise the analysis to meet the situation when some 
nodes are sending messages while others are not, we say that a node 
can be in state $S_1$ or $S_2$. In state $S_1$, 
a node can stabilise its first successor at rate $\alpha \lambda_s$, 
fail at rate $\lambda_f$ and assist in 
joins at rate $\lambda_j$ as before. In state $S_2$,
a node can stabilise its first successor at rate $a \lambda_s$, 
fail at rate $\lambda_f$, assist in joins at rate $\lambda_j$ 
and in addition, send correction messages 
(which is essentially equivalent to doing one lookup ) at 
rate $ \lambda_M \equiv  c \lambda_s$.  As we show in Section 
\ref{compare}, if we want to compare the two maintenance strategies
in a fair manner then the most general values that these
parameters can take is $\alpha=1$ and $a+c=1$.

Let $N_{S_1}$ be the number of nodes in state $S_1$ and 
$N_{S_2}$  the number of nodes in state $S_2$. Clearly
$N_{S_1} + N_{S_2} = N$, the total number of nodes in the system.

We can further partition $S_2$ into $S_{2}^{1}$,$S_{2}^{2}$,
$S_{2}^{3}$, $\cdots$, $S_{2}^{\cal M}$.
$S_{2}^{1}$ is the state of the node which has yet to send its first 
correction message, $S_{2}^{2}$ the state of the node which has 
sent its first correction message but is yet to send its second, {\it etc}.

Consider the gain and loss terms for $N_{S_1}$. These are summarised in 
table \ref{tab:ns1}.

\begin{table}
	\centering
		\begin{tabular}{|l|l|} \hline
		$N_{S_1}(t+\Delta t)$	&  Probability of Occurence   \\ %\hline 
		 $= N_{S_1}(t)-1$ & $c_{1.1}=(\lambda_f N_{S_1} \Delta t)$ \\ %\hline
		 $= N_{S_1}(t)+1$ & $c_{1.2}=(\lambda_j N \Delta t)$ \\ %\hline
		 $= N_{S_1}(t)+1$ & $c_{1.3}=(\lambda_M  N_{S_{2}^{\cal M}}\Delta t)$ \\ 
		 $= N_{S_1}-1$ & $c_{1.4}=(\alpha \lambda_s N_{S_1} \Delta t) w_1$\\ %\hline
		 $= N_{S_1}(t)$ & $1 - (c_{1.1} + c_{1.2} + c_{1.3} + c_{1.4})$\\ \hline
		\end{tabular}
\caption{Gain and loss terms for $N_{S_1}$ the number of nodes in state $S_1$.}
\label{tab:ns1}
\end{table}

Term $c_{1.1}$ is the probability that an $S_1$ node is lost because it 
failed. Term $c_{1.2}$ is the probability that a join occurs 
thus adding to the number of $S_1$ nodes in the system (since a new joinee
is always an $S_1$-type node). Term $c_{1.3}$ is the probability that
an $S_{2}^{\cal M}$ node sent its last message at rate $\lambda_{M}$ and
converted into an $S_1$ node. The last term $c_{1.4}$ is the probability
that an $S_1$-type node did a stabilisation at rate $\alpha \lambda_s$, 
found a wrong first successor with probability $w_1$ and hence 
converted into an $S_2$ node. $w_1$ is the fraction of wrong successor
pointers of an $S_1$-type node.

Defining $\lambda_s/\lambda_f = r$ and $\lambda_M/\lambda_f = cr$ 
the steady state equation predicted by table \ref{tab:ns1} is:
\begin{equation}
P_{S_1}(1+ \alpha rw_1) = 1+ cr P_{S_{2}^{\cal M}}
\label{eq:PS1}
\end{equation}

where $P_{S_1} = N_{S_1}/N$.

We can write a similar equation $N_{S_2}$ which however does not give 
us any new information since $N_{S_1} +N_{S_2} = N$.

Writing a gain-loss equation for each of the $N_{S_{2}^{i}}$'s in turn, 
we obtain,

\begin{equation}
P_{S_{2}^{1}} = \frac{P_{S_1}(\alpha rw_1 - arw_{1}^{\prime})}{1+cr+arw_{1}^{\prime}} + \frac{arw_{1}^{\prime}}{1+cr+arw_{1}^{\prime}}
\end{equation}

and

\begin{equation}
P_{S_{2}^{i}} = P_{S_{2}^{1}} \left(\frac{cr}{1+cr+arw_{1}^{\prime}} \right)^{i-1}
\end{equation},
for $2 \le i \le {\cal M}$.

Here $w_1$ is the fraction of $S_1$ nodes with wrong pointers and 
$w_{1}^{\prime}$ is the fraction of $S_2$ nodes with wrong pointers. 
We have made a simplification here in assuming that the fraction of wrong pointers
of $S_2$ nodes is the same, irrespective of the state of the $S_2$ node. In practice 
(especially if $a=0$),
this will not be the case. However
for the parameter ranges we are interested in ($r >>1$), this is not crucial.

Clearly $\sum_{1}^{\cal M} P_{S_{2}^{i}} = P_{S_2}$. A quantity of interest
in our analysis is

\begin{equation}
 P_{S_{2}^{\cal M}}/P_{S_2} = 1 - \frac{(1-g_1^{{\cal M}-1})}{1-g_1^{\cal M}}
\label{eq:order}
\end{equation}

%Using this, we can write the following equation:

%\begin{equation}
%1-P_{S_1} = rw_1P_{S_1} - g \left( arw_{1}^{\prime} + P_{S_1} \left( rw_1 - arw_{1}^{\prime} \right) \right) 
%\label{eq:ps1}
%\end{equation}

where $g_1= \frac{cr}{(1+cr+arw_{1}^{\prime})}$.

To solve for $P_{S_1}$ {\it etc}, we need to solve for 
$w_1$ and $w_{1}^{\prime}$.

However, consider first the equation for $W_T$ -- the {\it total} number of wrong
successor pointers in the system (irrespective of whether the pointer belongs
to an $S_1$ or an $S_2$ type node. The gain and loss terms for $W_T$ are shown in
table \ref{tab:wrong_total}. $w = W_T/N$ is the fraction of wrong succesor
pointers in the system.

\begin{table}[t]
\caption{Gain and loss terms for $W_T$: the total number of wrong first successor
pointers in the system.} 
\label{tab:wrong_total}
	\centering
		\begin{tabular}{|l|l|} \hline
		Change in $W_T$	&  \minorchange{Probability of Occurrence}   \\ %\hline 
		$W_T(t+\Delta t) = W_T(t)+1$ & $c_{2.1}=(\lambda_j N \Delta t) (1-w)$ \\ %\hline
		$W_T(t+\Delta t) = W_1(t)+1$ & $c_{2.2}=(\lambda_f N \Delta t) (1-w)^2 $ \\ %\hline
		$W_T(t+\Delta t) = W_1(t)-1$ & $c_{2.3}=(\lambda_f N \Delta t)$ \\ %\hline 
		$W_1(t+\Delta t) = W_1(t)-1$ & $c_{2.4}=(\alpha \lambda_s \Delta t) {N_{S_1}} w_1 + (a \lambda_s \Delta t) {N_{S_2}} w_{1}^{\prime}$\\ %\hline
		$W_1(t+\Delta t) = W_1(t)$ & $1 - (c_{2.1} + c_{2.2} + c_{2.3} + c_{2.4})$\\ 
\hline
		\end{tabular}
%\vspace*{-0.35cm}
\end{table}

This gives the following equation 

\begin{equation}
(3+\alpha r) w_1P_{S_1} + (3 +ar) w_{1}^{\prime}P_{S_2} =2
\label{eq:WT}
\end{equation}

The gain and loss terms $W_{1}^{\prime}$. 
-- the number of $S_2$ nodes with wrong successor pointers --
are written in much the same way 
except for a few small changes.
Table \ref{tab:wrong} details the changes that occur in $W_{1}^{\prime}$. 
in  time $\Delta t$.

\begin{table}[t]
\caption{Gain and loss terms for $W_{1}^{\prime}$: the number of wrong first successor
pointers of $S_2$-type nodes.} 
\label{tab:wrong}
	\centering
		\begin{tabular}{|l|l|} \hline
		Change in $W_1$	&  \minorchange{Probability of Occurrence}   \\ %\hline 
		$W_{1}^{\prime}(t+\Delta t) = W_{1}^{\prime}(t)+1$ & $c_{2.1}=(\lambda_j {N_{S_2}} \Delta t) (1- w_{1}^{\prime})$.  \\ %\hline
		$W_{1}^{\prime}(t+\Delta t) = W_{1}^{\prime}(t)+1$ & $c_{2.2}=\lambda_f {N_{S_2}} (1-w_{1}^{\prime})^2 P_{S_2} $ \\ %\hline
& $+ (1-w_1)(1-w_{1}^{\prime}) P_{S_1})   \Delta t$ \\ %\hline
		$W_{1}^{\prime}(t+\Delta t) = W_{1}^{\prime}(t)-1$ & $c_{2.3}=\lambda_f {N_{S_{2}}} {(w_{1}^{\prime}}^2 P_{S_2} + w_1w_{1}^{\prime} P_{S_1})   \Delta t $ \\ 
		$W_{1}^{\prime}(t+\Delta t) = W_{1}^{\prime}(t)-1$ & $c_{2.4}= a \lambda_s {N_{S_2}} w_{1}^{\prime}   \Delta t $\\ %\hline
$W_{1}^{\prime}(t+\Delta t) = W_{1}^{\prime}(t)-1$ & $c_{2.5}=\lambda_M {N_{S_2}^{\cal M}} w_{1}^{\prime}   \Delta t $\\ %\hline
		$W_1(t+\Delta t) = W_1(t)$ & $1 - (c_{2.1} + c_{2.2} + c_{2.3} + c_{2.4} +c_{2.5})$\\ 
\hline
		\end{tabular}
%\vspace*{-0.35cm}
\end{table}

The terms here are much the same as derived earlier except that
we now have to keep track of whether the node that is failing
(in terms $c_{2.2}$ and  $c_{2.3}$) is a $S_1$ or an $S_2$-type node.
In addition term $c_{2.5}$ is the probability that an $S_{2}^{\cal M}$-type
node has a wrong successor pointer, but sends a message and hence 
turns into an $S_1$ node with a wrong pointer.

Table \ref{tab:wrong} gives us the following equation for
$w_{1}^{\prime}$ in the steady state

\begin{equation}
2= w_{1}^{\prime} \left(3+a r + cr\frac{P_{S_{2}^{\cal M}}}{P_{S_2}} \right)
+( w_1 - w_{1}^{\prime})P_{S_1}
\label{eq:Wprime}
\end{equation}

We can write a similar equation for $w_{1}$ which however does not contain 
any new information since $w_1$ and $w_{1}^{\prime}$ satisfy equation
\ref{eq:WT}.

So in effect we have three equations, Eqn. \ref{eq:PS1}, Eq. \ref{eq:WT} and 
\ref{eq:Wprime} for three unknowns $P_{S_1}$, $w_1$ and $w_{1}^{\prime}$.  
In practice this set of equations is very hard to solve exactly because of the
appearance of terms such as  $g_1^{\cal M}$ in Eq. \ref{eq:order}.

In the following we will solve the set of equation to $O(1/r)$ by expanding
Eq. \ref{eq:order} to first order in $w_{1}^{\prime}$.  In this case, 

\begin{equation}
 P_{S_{2}^{\cal M}}/P_{S_2} = \frac{1}{\cal M} - \left(\frac{{\cal M}-1}{2 {\cal M}}\right)\frac{1+arw_{1}^{\prime}}{cr}
\label{eq:order1byr}
\end{equation}

We can now solve the set of three coupled equations to get a quartic equation for
$w_{1}^{\prime}$ as a function of $a,\alpha,{\cal M} $ and $r$. Only one of the
roots of the quartic equation is a true solution satisfying all the conditions above.
The details of the calculations  though straight forward are tedious and not shown here.

To calculate the cost of lookups, we still need to calculate the
probability that a finger is dead.  The loss and gain terms for this calculation are 
almost exactly  the same as  carried out earlier, in \cite{KEAH1,KEAH2} 
(except for term $c_{3.2}$) and are shown in table \ref{tab:f}.  

\begin{table}
\caption{The relevant gain and loss terms for $F_k$, the number of nodes whose $k{th}$ fingers are pointing to a failed node for $k > 1$.}
\label{tab:f}
	\centering
		\begin{tabular}{|l|l|} \hline
		$F_k(t+\Delta t)$	&  \minorchange{Probability of Occurence}  \\ %\hline 
		$= F_k(t)+1$ & $c_{3.1}=(\lambda_j{N} \Delta t) \sum_{i=1}^{k}p_{\minorchange{\it join}}(i,k)f_i$
		\\ %\hline
		$= F_k(t)-1$ & $c_{3.2}= \frac{f_k}{\sum_k f_k} (\lambda_M {N_{S_2}}(1- w_{1}^{\prime}) A(w_1,w_{1}^{\prime}) \Delta t)$ \\ %\hline
		$= F_k(t)+1$ & $c_{3.3}= (1-f_k)^2 [1-p_{1}(k)] (\lambda_f {N} \Delta t)$ \\ %\hline
		$= F_k(t)+2$ & $c_{3.4}= (1-f_k)^2 (p_{1}(k)-p_{2}(k)) (\lambda_f{N} \Delta t)$ \\ %\hline
	  $= F_k(t)+3$ & $c_{3.5}= (1-f_k)^2 (p_{2}(k)-p_{3}(k)) (\lambda_f {N} \Delta t)$ \\ %\hline
		$= F_k(t)$   & $1 - (c_{3.1} + c_{3.2} + c_{3.3}+ c_{3.4}+ c_{3.5})$\\ \hline		
		\end{tabular}
\end{table}

The term $c_{3.2}$ is the probability that a message is sent ($\lambda_M N_{S_2}$) 
times the probability that a $k^{th}$ pointer gets  
this message (with probability 
$f_k/\sum f_k$ since {\it only} nodes with wrong pointers 
get the messages), times the probability that the message is not outdated
($1-w_{1}^{\prime}$), times the probability that the predecessor of the node
which has to receive the message has a correct successor pointer. This
last quantity is denoted by 
$A(w_1,w_{1}^{\prime})= 1- (w_1 P_{S_1} + w_{1}^{\prime} P_ {S_2})$, since the predecessor
could have been an $S_1$ or an $S_2$ type node.

An estimate for $\sum f_k$ is simply $\sim {\cal M} N_{S_2}/N$. Substituting this
in term $c_{3.2}$, this term becomes $= \lambda_M N \Delta t (f_k/{\cal M}) (1-w_{1}^{\prime}) A(w_1,w_{1}^{\prime})$

Solving for $f_k$ in the steady state, and substituting for  $w_{1}^{\prime}$, we
get $f_k$ as a function of the parameters. As mentioned earlier a  quick and precise 
estimate of the lookup length is then
obtained by taking $L= A(1+f+3f^2)$.

\subsection{Comparison of Correction-on-change and Periodic Stabilisation}
\label{compare}

In order to compare how the two strategies perform
under churn, we need to make sure that
we are comparing lookup latencies for the same number of total maintenance
messages sent.

Let us assume that the maximum rate for sending messages per node is $C$.
In the case of periodic stabilisation, this implies
that the rate of doing successor stabilisations $\lambda_{s_1}$ and finger
stabilisations $\lambda_{s_2}$ must in total not exceeed $C$.
This implies that $\lambda_{s_1}/C + \lambda_{s_2}/C \le 1$. 
If we assume that all nodes always send messages up to their maximum capacity, then
clearly $\lambda_{s_1}/C + \lambda_{s_2}/C = 1$. 
Suppose we define $r\equiv C/\lambda_j$ and $r_1 \equiv \lambda_{s_1}/ \lambda_j, r_2 \equiv \lambda_{s_2}/\lambda_j$. Then  for a given value of 
$r$, $r_1+r_2=r$. Hence if finger 
stabilisations are done at rate $(1-\beta) r$, the successor stabilisations 
need to be done at rate $\beta r$, where
the parameter $\beta$ can be varied from $0$ to $1$.

In the case of correction-on-change, we need to impose the same 
maximum rate $C$
no matter which state the nodes are in.
In this case,  let
$\lambda_{S_1}$ be the rate of successor stabilisation in state $S_1$,
$\lambda_{S_2}$ the rate of successor stabilisation in state $S_2$ and
$\lambda_{S_3}$ be the rate of sending messages in state $S_2$.
Clearly $\lambda_{S_1}=C$ and $\lambda_{S_2} + \lambda_{S_3} =C$.
Defining $r$ as before, we get
$\lambda_{s_1}/ \lambda_j = r$ and 
$\lambda_{s_2}/ \lambda_j + \lambda_{s_3}/\lambda_j = r$.
Hence comparing with our parameters
$\alpha=1$ and $a+c=1$.

\begin{figure}[t]
	\centering
	\includegraphics[height=8cm, angle=270]{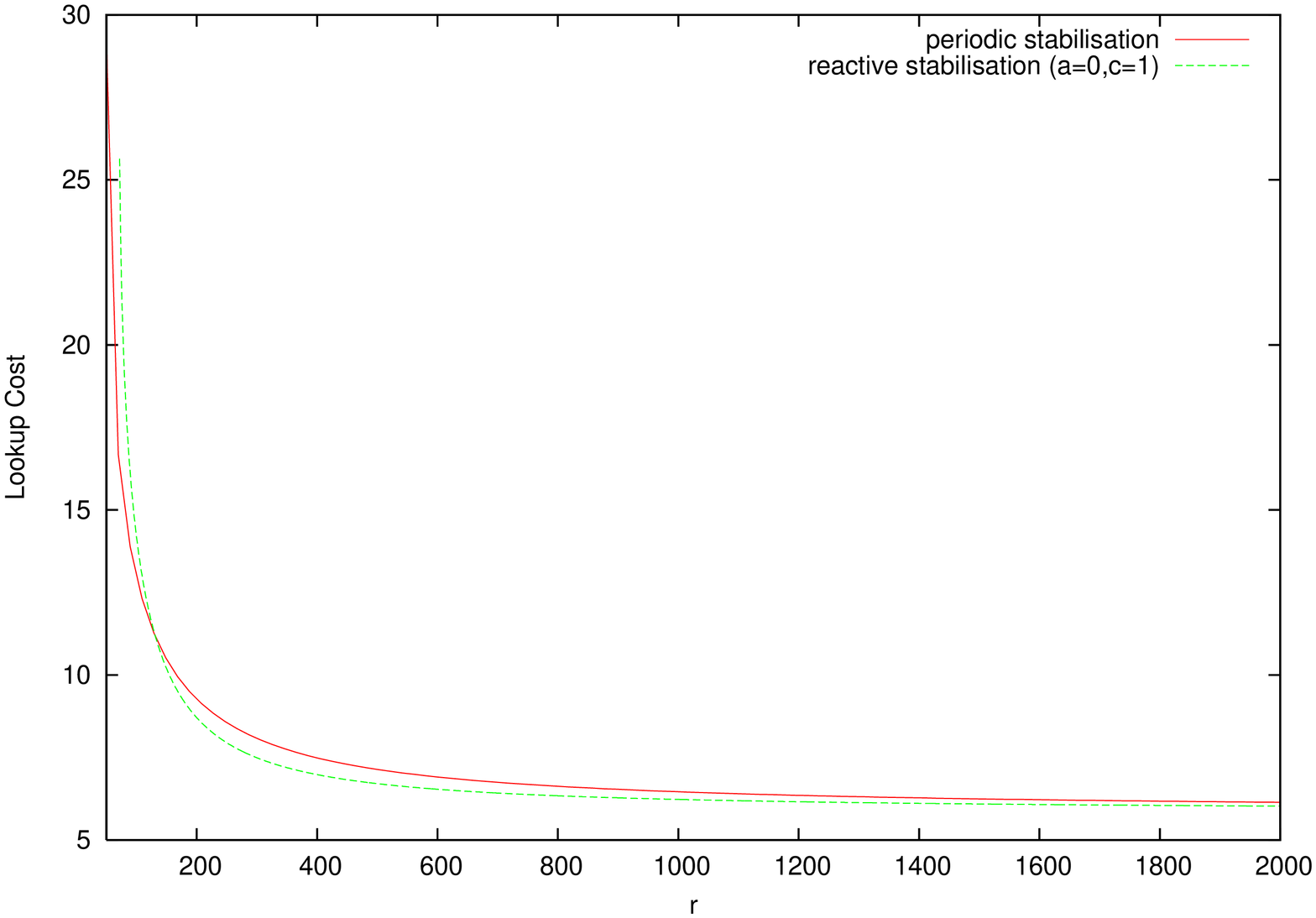}
	\caption{Comparison of the Lookup cost for the two maintenance 
strategies, for $N=1000$.}
	\label{fig:coc}
\end{figure}

In Fig. ~\ref{fig:coc}, we have plotted the function
$L= A(1+f+3f^2)$  with the value of the lookup length 
without churn $A=5.846$ for $N=1000$ nodes, 
for $a=0$ (and $c=1$) and
for $\beta=0.4$. $f$ is calculated separately for the two maintenance techniques.

\begin{figure}[t]
	\centering
	\includegraphics[height=8cm, angle=270]{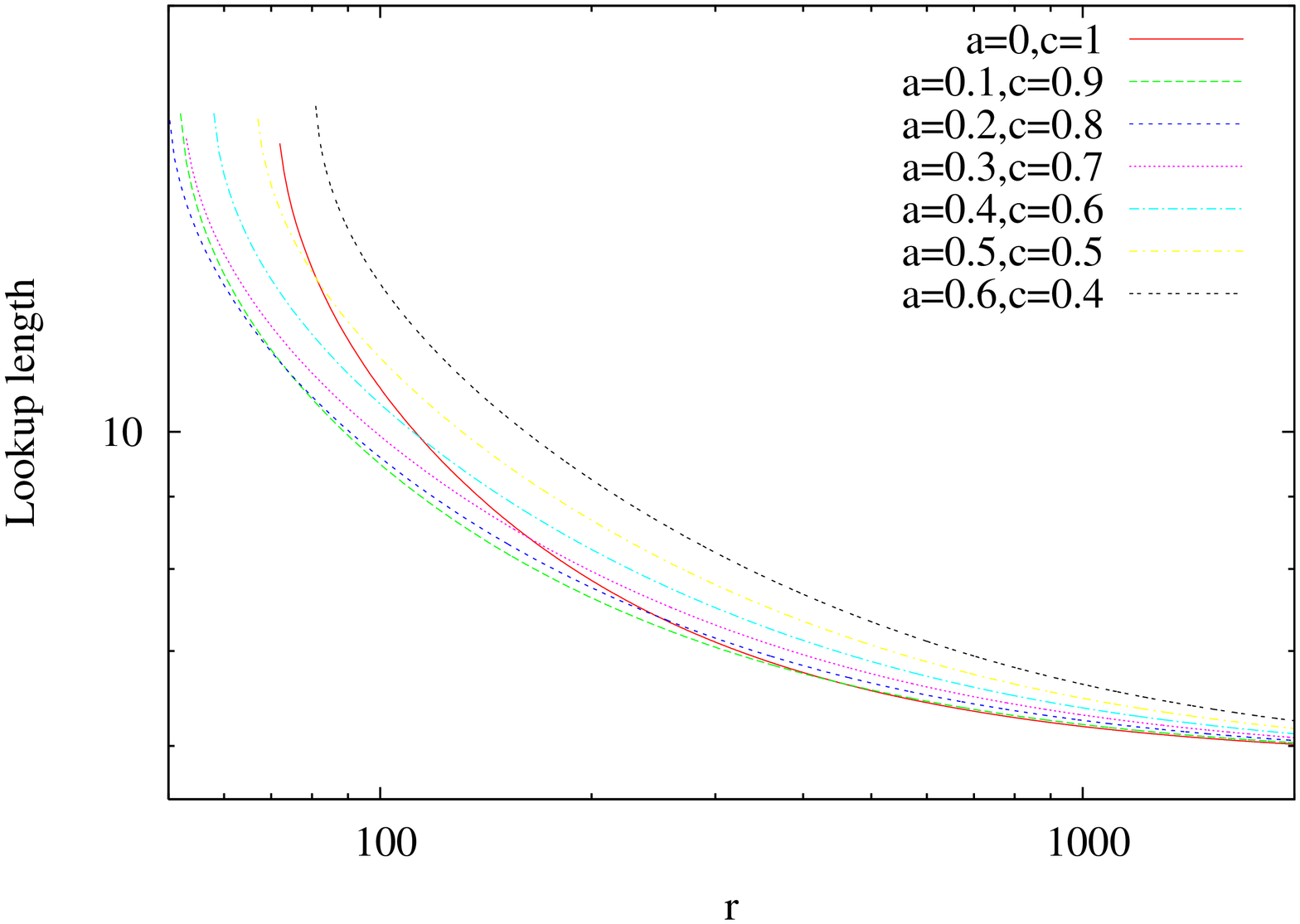}
	\caption{Comparison of the Lookup cost for different values of the parameter $a$, as explained in the text.}
	\label{fig:coc1}
\end{figure}

As can be seen, correction-on-change is better than periodic stabilisation
when churn is low but not when churn is high. On
comparing lookup lengths for several different $a$, it becomes evident 
(see yFig. ~\ref{fig:coc1}) that
$a\sim 0.2$ is the optimum value for the correction-on-change strategy.

So interestingly, for nodes in state $S_2$, it is not the
best strategy to increase $c$ as much as possible. Its a better strategy
to spend some of the bandwidth on maintaining a correct successor. 

\section{Summary}
\label{summary}
In summary, we have demonstrated the usefulness of the 
master-equation approach for understanding churn in overlay networks.
Our analysis can take into account most details of the
algorithms used by these networks, to provide predictions for
how the performance depends on the parameters. 
There are several directions in which we can extend the present analysis.
One of the more important ones is to model congestion on the links.
This could affect the performance of the two compared 
maintenance strategies differently. The periodic case
may not be as affected as much as the reactive case, which could suffer from 
congestion collapse.

{\bf Acknowledgments} We would like to thank Ali Ghodsi for several very useful discussions.

{\footnotesize
\bibliographystyle{amsplain}
\bibliography{P2P_v2}
}

\section{Appendix}
\label{A1}

Equation \ref{eq:cost} with the churn-dependent terms set to zero becomes:

\begin{equation}
\label{eq:costnochurn}
C_{\xi+m} = C_{\xi} \left[1-a(m)\right] + a(m) + \sum_{i=0}^{m-1}b(i)C_{m-i}
\end{equation}

After some rewriting of this, it is easily seen that the cost for {\it any}
key $i+1$ can be written as the following recursion relation:

\begin{equation}
\label{eq:constnochurn1}
C_{i+1} = \rho C_i + (1-\rho) +(1-\rho)C_{i+1-\xi(i+1)}
\end{equation}

Here we have used the definition of $a$ and $b$ from the internode-interval distribution and the notation $\xi(i+1)$ refers to the {\it start} of the 
finger most closely preceding $i+1$. For instance, for $i+1=4$, 
$\xi(i+1)=2$ and for $i+1=11$, $\xi(i+1)=8$ etc.

\begin{figure}[t]
	\centering
	\includegraphics[height=8cm, angle=270]{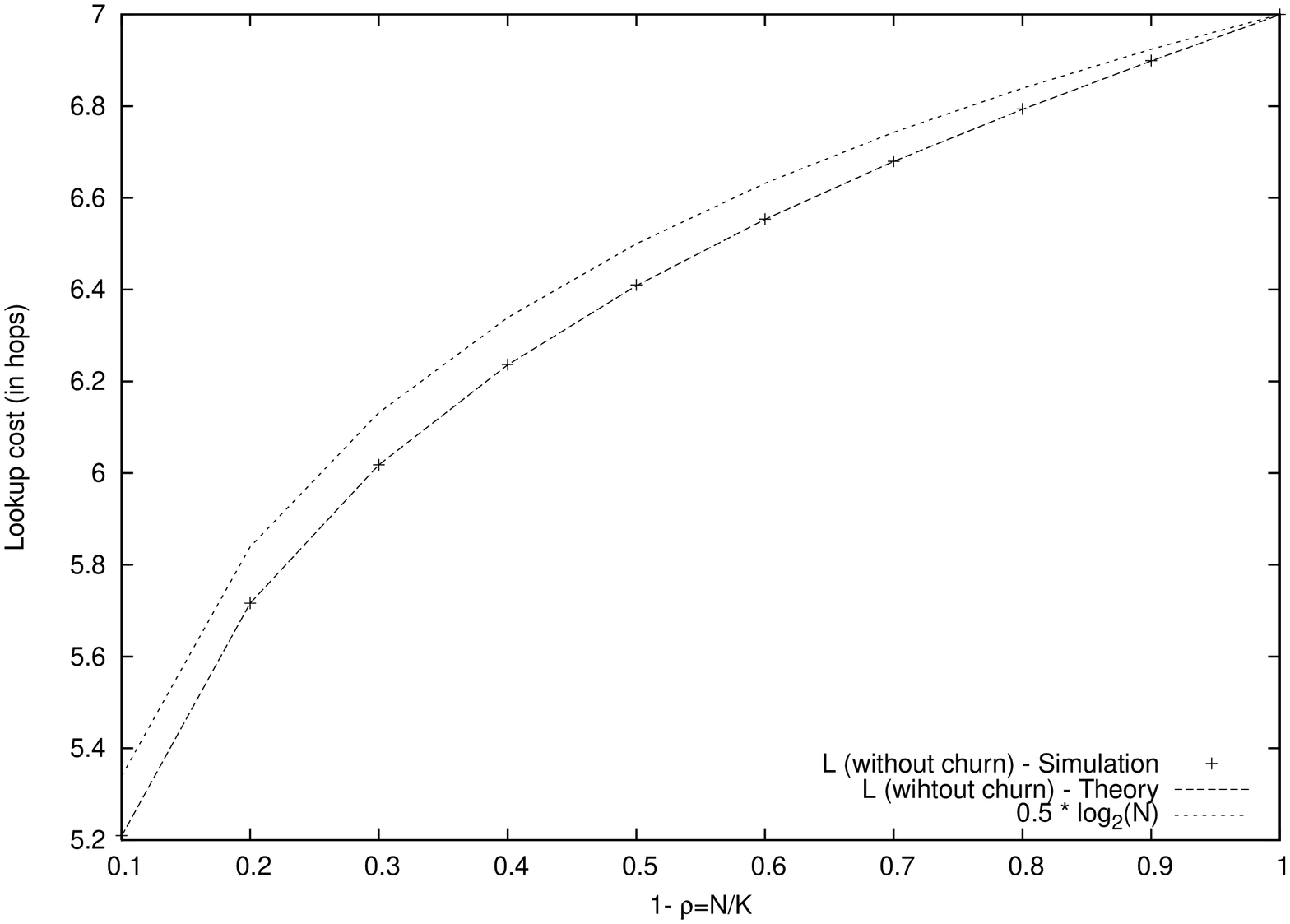}
	\caption{Theory and Simulation for the lookup cost without churn for a key space of size ${\cal K}=2^{14}$ for varying $N$. Plotted as reference is the curve $0.5 \log_2(N)$. Note that on the y axis we have actually plotted 
$L-1$ for convenience.}
	\label{fig:nochurn}
\end{figure}

\begin{figure}[t]
	\centering
	\includegraphics[height=8cm, angle=270]{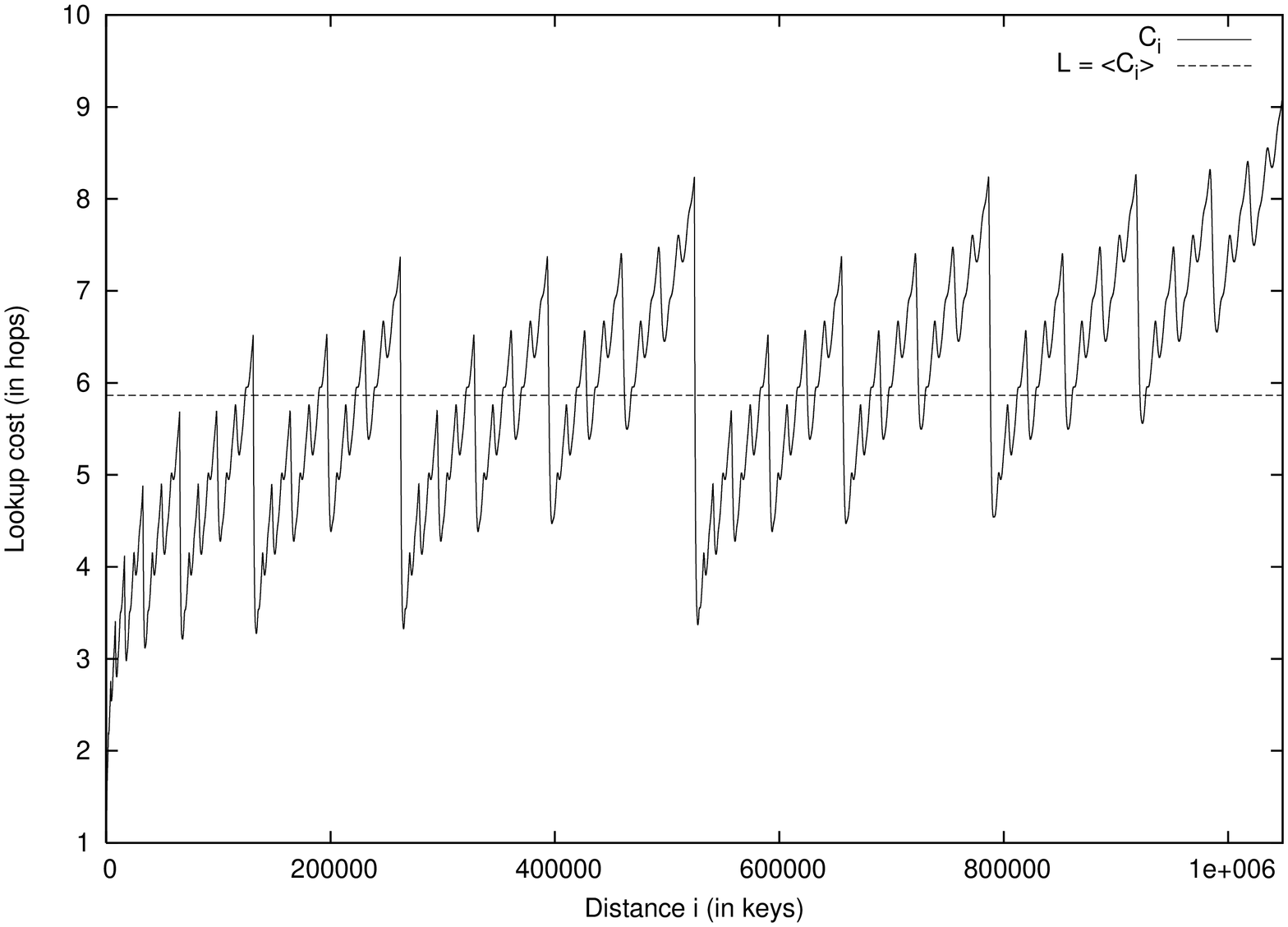}
	\caption{The average cost $C_i$ (the number hops for looking up an 
item $i$ keys away) in a network of ${\cal N}=1000$ nodes and 
${\cal K}= 2^{20} $ keys without churn obtained from the 
recurrence relation (\ref{eq:constnochurn1}). The average lookup length $L$ is also plotted as a reference.}
	\label{fig:ici}
\end{figure}

We are interested in solving the recursion relation and computing  $L= \frac{1}{{\cal K}}\sum_{i=1}^{{\cal K}-1}C_i$.
To do this, we decompose this sum into the following partial sums:
\begin{equation}
\begin{split}
s_0 &= C_1 = 1    \\
s_1 &= C_2        \\
s_2 &= C_3 + C_4  \\
s_3 &= C_5 + C_6 + C_7 + C_8 \\
\ldots \\
s_{\cal M} &= C_{2^{{\cal M}-1}+1}+\ldots+C_{{\cal K}-1}
\end{split}
\end{equation}
Substituting the expressions for the $C$'s in the above, we find:
\begin{equation}
\begin{split}
s_0 &= 1 \\
s_1 &= \frac{\rho}{1-\rho}[C_1-C_2]+ 1 +  s_0 \\
s_2 &= \frac{\rho}{1-\rho}[C_2-C_4]+ 2 + [s_0+s_1] \\
\ldots \\
s_i &= \frac{\rho}{1-\rho}[C_{2^{i-1}}-C_{2^i}]+ 2^{i-1} + \sum_{j=0}^{j-1} s_j
\end{split}
\end{equation}
By substituting serially the expressions for $s_j$ (where $0 \leq j \leq i-1$), the expression for $s_i$
 (for $i \geq 2$) becomes:
\begin{equation}
\begin{split}
s_i &= \frac{\rho}{1-\rho} [2^{i-2}C_1 - C_{2^i} - \sum_{j=1}^{i-2} s^{i-2-j}C_{2^j}] \\
   &+ 2^{i}+ (i-1)2^{i-2} 
\end{split}
\end{equation}
Hence
\begin{equation}
\begin{split}
\sum_{i=0}^{\cal M} s_i & = -\rho + [2^{{\cal M}+1}-1] + {\cal M}2^{{\cal M}-1} - [2^{{\cal M}}-1]  \\
&+ \frac{\rho}{1-\rho} \biggl[(2^{{\cal M}-1}-1)C_1 - \sum_{i=2}^{{\cal}M -1}C_{2^i}  - C_{{\cal K}-1}  \\
&- (2^{{\cal M}-2}-1)C_2 -  (2^{{\cal M}-3}-1)C_4 - \dots \biggr]
\end{split}
\end{equation}
Therefore
\begin{equation}
\begin{split}
\sum_{i=0}^{\cal M} s_i & = -\rho + 2^{\cal M} + {\cal M}2^{{\cal M}-1} \\
&+ \frac{\rho}{1-\rho} \biggl[(2^{{\cal M}-1}-1)C_1 - \sum_{i=2}^{{\cal}M -1}C_{2^i}  - C_{{\cal K}-1}  \\
-& \sum_{j=2}^{{\cal M}-2} (2^{{\cal M}-j}-1)C_{2^{j-1}}  \biggr]
\end{split}
\end{equation}
The equation for the average lookup length without churn is thus,
\begin{equation}
\begin{split}
	L &= \frac{\sum s}{\cal K} \\
	  &= -\frac{\rho}{\cal K} +1 + \frac{1}{2}{\cal M} \\
	  &+ \frac{\rho}{1-\rho} \biggl[ \frac{2^{{\cal M}-1}-1}{\cal K}C_1 
	                               - \frac{1}{\cal K} \sum_{i=2}^{{\cal M}-1}C_{2^i}
	                               - \frac{1}{\cal K} C_{{\cal K} -1} \\
	                              -& \sum_{j=2}^{{\cal M}-2} \frac{2^{{\cal M}-j}-1}{\cal K} C_{2^{j-1}} \biggr]
\end{split}
\end{equation}

If we can take the limit ${\cal K} \to \infty$, we can throw away some of the terms.
\begin{equation}
\begin{split}
\lim_{{\cal K}\to\infty}	L &= 1 + \frac{1}{2}{\cal M} \\
	  &+ \frac{\rho}{1-\rho} \biggl[\frac{C_1}{2}  
	                               - \frac{1}{\cal K} \sum_{i=1}^{{\cal M}-1}C_{2^i} 
	                               +\frac{C_2}{\cal K}
	                               - \frac{1}{\cal K} C_{{\cal K} -1} \\
	                               -& \sum_{j=2}^{{\cal M}-2} \frac{2^{{\cal M}-j}}{\cal K} C_{2^{j-1}} 
	                               + \sum_{j=2}^{{\cal M}-2} \frac{C_{2^{j-1}}}{\cal K} \biggr]\\ 
	                               \approx & 1 + \frac{1}{2}{\cal M} + \frac{\rho}{1-\rho}
	                               \left[\frac{C_1}{2}- \frac{C_2}{4}- \frac{C_4}{8} \ldots  
	                               - \frac{C_{2^{{\cal M}-3}}}{2^{{\cal M}-2}}\right] 
\end{split}
\end{equation}
Since $C_1 = 1$, we can write 
\begin{equation}
\label{eq:Kinf}
\begin{split}
    L &= 1 + \frac{1}{2}{\cal M} - \frac{\rho}{2(1-\rho)}
	                               \biggl[\frac{C_2 -1}{2} + \frac{C_4 -1 }{4}+  \ldots \\
	                               +& \frac{C_{2^{{\cal M}-3}}-1}{2^{{\cal M}-3}}\biggr] 
\end{split}
\end{equation}
From the recursion relation for the $C_i$'s, it is easy to see that 
\begin{equation}
(C_i-1) = (1-\rho) g_i^{(1)}(\rho)+(1-\rho)^2 g_i^{(2)}(\rho) + \ldots
\end{equation}
where the $g_i$'s are functions only of $\rho$.

Hence if ($1-\rho$) is small ($\frac{N}{\cal K}\to 0$), we need only compute the $C_i$'s to
first order in ($1-\rho$) to get the leading order effect and second order in ($1-\rho$) to get the correction etc.

Hence in general the, the expression for $L$ is:
\begin{equation}
\label{eq:exact}
\begin{split}
    L &= 1 + \frac{1}{2}{\cal M} - \frac{\rho}{2} 
	                               \biggl[e_1(\rho)+(1-\rho)e_2(\rho)+(1-\rho)^2 e_3(\rho) \ldots \biggr] 
\end{split}
\end{equation}
Where $e_1(\rho)=\sum_{i=1}^{{\cal M}-3} g_{2^i}^{(1)}(\rho)$ etc.

We evaluate this expression numerically by solving recursion relation (\ref{eq:constnochurn1}) and compare it with 
simulations done at zero churn. As can be seen the prediction of the equation is very accurate (Figure \ref{fig:nochurn}).

Let us now compute $e_1(\rho)$ to see what the leading order effect is. We now need to solve recursion relation (\ref{eq:constnochurn1}) only to order $1-\rho$, which gives:
\begin{equation}
\begin{split}
	C_2-1 &= (1-\rho) \\
	C_4-1 &= (1-\rho)\left[1+ \rho +\rho^2 \right] \\
	C_8-1 &= (1-\rho)\left[1+ \rho +\rho^2+ \dots + \rho^6 \right] \\
	\ldots\\
	C_i-1 &= (1-\rho)\left[1+ \rho +\rho^2+ \dots + \rho^{i-2} \right] \\
\end{split}
\end{equation}
Therefore,
\begin{equation}
\begin{split}
    L &= 1 + \frac{1}{2}{\cal M} + \frac{\rho}{2} 
	                               \biggl[\frac{1}{2}+\frac{1+\rho+\rho^2}{4}+ \ldots \biggr] 
\end{split}
\end{equation}
Consider the expression inside the brackets. We are computing this in the approximation $\frac{N}{\cal K}=\epsilon \to 0$, i.e. $\rho =1 - \epsilon$, therefore $\rho^x = (1-\epsilon)^x \approx e^{-\epsilon x}$. If $x > \frac{1}{\epsilon}$, then $\rho^x \to 0$, therefore if $x > \frac{\cal K}{N}$, then $\rho^x \to 0$. Hence, the terms inside the brackets become:
\begin{equation}
\label{eq:T}
\begin{split}
\sum_{j=1}^{T} \frac{2^{j}-1}{2^j} + (2^T -1)\sum_{j=T+1}^{{\cal M}-3} \frac{1}{2}j
\end{split}
\end{equation}
Where $T \equiv \ln_2{\cal K}-\ln_2{N}$ and we have put $\rho^x \approx 1$ for $x < \frac{\cal K}{N}$ and $\rho \to 0$ for $x > \frac{\cal K}{N}$. This is clearly an overestimation and so we expect the result to over estimate the exact expression
\ref{eq:exact}. 

Expression \ref{eq:T} becomes:

$$T-\left[1-(\frac{1}{2})^{{\cal M}-3}\right] + \left[1-(\frac{1}{2})^{{\cal M}-3-T}\right] \approx T$$
Therefore:
\begin{equation}
\label{eq:chder}
\begin{split}
L & = 1+ \frac{1}{2}{\ln_2{\cal K}} - \frac{1}{2}\left[ \ln_2{\cal K} - \ln_2N\right] \\
  & \approx 1+ \frac{1}{2}\ln_2N 
\end{split}
\end{equation}
Which is the known result for the average lookup length of Chord.

Another important parameter in the performance of DHTs in general is the
base. By increasing the base, the number of fingers per node increases
which leads to a shorter lookup path length. The effect of varying the
base has been studied in \cite{onana03dks,dhtcomparison:infocom05}. So
far, we have considered in this analysis base-$2$ Chord. We can likewise
carry out this analysis for any base.

In general, we have base-$b$ with $(b-1) log_b({\cal K})$ fingers per  node. 
Consider as an example $b=4$. Here we can define the
the partial sums again in the following manner:\\
\begin{equation}
\begin{split}
	\Delta_0 &= s_0 = C_1=1 \\
	\Delta_1 &= s_1 + s_2 + s_3 \\
	\Delta_2 &= s_4 + s_5 + s_6 \\
         \ldots & 
\end{split}
\end{equation}
where 
\begin{equation}
\begin{split}
	s_1 &= C_2 = \rho C_1 + (1-\rho) + (1-\rho)C_1\\
	s_2 &= C_3 = \rho C_2 + (1-\rho) + (1-\rho)C_1\\
	s_3 &= C_4 = \rho C_3 + (1-\rho) + (1-\rho)C_1\\
	s_4 &= C_5 + C_6 + C_7 + C_8 \\
	s_5 &= C_9 + C_{10} + C_{11} + C_{12} \\
	s_6 &= C_{13} + C_{14} + C_{15} + C_{16}\\
	\ldots & 
\end{split}
\end{equation}
Therefore
\begin{equation}
\begin{split}
	\Delta_0 &= C_1 \\
	\Delta_1 &= \rho \left[\Delta_1 + C_1 -C_4  \right] + 3 (1-\rho) + 3 (1-\rho) \left[ \Delta_0 \right] \\
	\Delta_2 &= \rho \left[\Delta_2 + C_4 -C_{16} \right] + 12(1-\rho) + 3 (1-\rho) \left[ \Delta_0+\Delta_1 \right] \\
	\ldots & 
\end{split}
\end{equation}
In general for a base $b$, define $B \equiv b-1$ and $b^{\cal M}={\cal K}$.
Then we have:
\begin{equation}
\begin{split}
	\Delta_j &= \frac{\rho}{1-\rho} \left[C_{b^{j-1}} - C_{b^j} \right] \\
	+& B(B+1)^{j-1} + B \left[ \Delta_0+ \Delta_1+ \dots + \Delta_{j-1} \right]
\end{split}
\end{equation}

Following much the same procedure as before, we find
\begin{equation}
\begin{split}
L =& \frac{1}{\cal K} \sum_{j=0}^{\cal M}\Delta_j \\
\approx & 1+ \frac{B}{B+1} {\cal M} - \frac{B}{B+1} \frac{\rho}{1-\rho} \left[ \frac{C_b-1}{B+1} + \frac{C_{b^2}-1}{(B+1)^2} + \ldots \right]
\end{split}
\end{equation}
for ${\cal K} \to \infty$ as the analogue of (\ref{eq:Kinf}).
Again we can simplify and slightly overestimate the sum by assuming that $\rho^x \approx 0$ for $x > \frac{\cal K}{N}$ and $\rho^x \approx 1$ for $x < \frac{\cal K}{N}$. Then we get:
\begin{equation}
\begin{split}
L \approx 1 + \frac{b-1}{b}\frac{\ln_2N}{\ln_2b}
\end{split}
\end{equation}

This is the analogue of Eq. \ref{eq:chder} for any base $b$.

%\end{multicols}
%\clearpage
%\input{fixsucc.tex}
%\input{fixfingers.tex}

{\footnotesize
\bibliographystyle{amsplain}
\bibliography{P2P_v2}
}
%\end{multicols}
%\clearpage
%\input{fixsucc.tex}
%\input{fixfingers.tex}

\end{document}